\documentclass{aastex631}

\usepackage{savesym}
\usepackage{lineno}
\usepackage[sort&compress]{natbib}

\usepackage{graphicx} 
\usepackage{gensymb}
\modulolinenumbers[1]

\usepackage{booktabs}
\savesymbol{splitbox}
\usepackage{adjustbox}

\graphicspath{{./}{}}

\newcommand{\source}[1]{\textsuperscript{\textcolor{blue}{[citation needed]}}\xspace}

\shorttitle{Arpu Kuilpu}
\shortauthors{Shober et al.}

\begin{document}
\title{Arpu Kuilpu: An H5 from the Outer Main Belt}

\correspondingauthor{Patrick M. Shober}
\email{patrick.shober@postgrad.curtin.edu.au}

\author[0000-0003-4766-2098]{Patrick M. Shober}
\affiliation{Space Science \& Technology Centre, School of Earth and Planetary Sciences, Curtin University, Bentley, WA 6102, Australia}

\author[0000-0001-9226-1870]{Hadrien A. R. Devillepoix}
\affiliation{Space Science \& Technology Centre, School of Earth and Planetary Sciences, Curtin University, Bentley, WA 6102, Australia}

\author[0000-0003-2702-673X]{Eleanor K. Sansom}
\affiliation{Space Science \& Technology Centre, School of Earth and Planetary Sciences, Curtin University, Bentley, WA 6102, Australia}

\author[0000-0002-8240-4150]{Martin C. Towner}
\affiliation{Space Science \& Technology Centre, School of Earth and Planetary Sciences, Curtin University, Bentley, WA 6102, Australia}

\author[0000-0003-2193-0867]{Martin Cup\'ak}
\affiliation{Space Science \& Technology Centre, School of Earth and Planetary Sciences, Curtin University, Bentley, WA 6102, Australia}

\author[0000-0002-8914-3264]{Seamus L. Anderson}
\affiliation{Space Science \& Technology Centre, School of Earth and Planetary Sciences, Curtin University, Bentley, WA 6102, Australia}

\author[0000-0003-0990-8878]{Gretchen Benedix}
\affiliation{Space Science \& Technology Centre, School of Earth and Planetary Sciences, Curtin University, Bentley, WA 6102, Australia}

\author[0000-0002-5759-6517]{Lucy Forman}
\affiliation{Space Science \& Technology Centre, School of Earth and Planetary Sciences, Curtin University, Bentley, WA 6102, Australia}

\author[0000-0002-4681-7898]{Phil A. Bland}
\affiliation{Space Science \& Technology Centre, School of Earth and Planetary Sciences, Curtin University, Bentley, WA 6102, Australia}

\author[0000-0002-5864-105X]{Robert M. Howie}
\affiliation{Space Science \& Technology Centre, School of Earth and Planetary Sciences, Curtin University, Bentley, WA 6102, Australia}

\author[0000-0002-8646-0635]{Benjamin A. D. Hartig}
\affiliation{Space Science \& Technology Centre, School of Earth and Planetary Sciences, Curtin University, Bentley, WA 6102, Australia}

\author[0000-0001-5390-4343]{Matthias Laubenstein}
\affiliation{Laboratori Nazionali del Gran Sasso, Ist. Nazionale di Fisica Nucleare, Via G. Acitelli 22, I-67100 Assergi (AQ), Italy}

 \author[0000-0001-9217-0960]{Francesca Cary}
 \affiliation{Space Science \& Technology Centre, School of Earth and Planetary Sciences, Curtin University, Bentley, WA 6102, Australia}

\author[0000-0002-4394-9721]{Andrew Langendam}
 \affiliation{Department of Earth and Environmental Sciences, Macquarie University, North Ryde, Sydney, NSW, 2109, Australia}

\begin{abstract}
On 1 June 2019, just before 7:30\,PM local time, the Desert Fireball Network detected a -9.3 magnitude fireball over South Australia near the Western Australia border. The event was observed by six fireball observatories, and lasted for five seconds. One station was nearly directly underneath the trajectory, greatly constraining the trajectory solution. This trajectory's backward numerical integrations indicate that the object originated from the outer main belt with a semi-major axis of 2.75~au. A light curve was also extracted and showed that the body experienced very little fragmentation during its atmospheric passage. A search campaign was conducted with several Desert Fireball Network team members and other volunteers. One 42~g fragment was recovered within the predicted fall area based on the dark flight model. Based on measurements of short-lived radionuclides, the fragment was confirmed to be a fresh fall. The meteorite, Arpu Kuilpu, has been classified as an H5 ordinary chondrite. This marks the fifth fall recovered in Australia by the Desert Fireball Network, and the smallest meteoroid ($\simeq 2\,kg$) to ever survive entry and be recovered as a meteorite.
\end{abstract}

\section{Introduction} \label{sec:intro}
The Desert Fireball Network (DFN) is a system of automated photographic all-sky fireball observatories covering over 2.5 million km$^{2}$ of Australian outback \citep{Howie2017Howbuildcontinental}. It is the largest single photographic fireball network in the world. The DFN's primary objective is to recover meteorite falls using fireball observations. These meteorites with precise orbital information can help us better understand small bodies and the debris they generate in the solar system. The DFN is a partner network of the Global Fireball Observatory (GFO), a collaborative project consisting of 18 international partners from all over the world \citep{devillepoix2020global}. Our observatories are optimized to precisely observe bright meteors (``fireballs'') when they impact the atmosphere, providing invaluable information about the asteroidal debris in the inner solar system \citep{Devillepoix2019Observationmetrescale}. 

Over forty meteorite falls in total have been recovered with the assistance of fireball observations \citep{granvik2018identification,devillepoix2020global,colas2020fripon,gardiol2021cavezzo}.
The most numerous of the meteorite fall types recovered are H-chondrites.  This is somewhat unexpected, as the most common Antarctic meteorite type are L-chondrites, followed by H-chondrites \citep{binzel2015near}. More meteorite fall recoveries can clarify whether this trend is significant. Within this study, we discuss the DFN fireball observations and data reduction, which lead to the recovery of Arpu Kuilpu\footnote{\url{https://www.lpi.usra.edu/meteor/metbull.php?code=74013}}, an H5 chondrite. This is the fifth meteorite recovered in Australia by the DFN \citep{Bland2009AnomalousBasalticMeteorite,Devillepoix2018DingleDellmeteorite,Dyl2016CharacterizationMasonGully,sansom2020murrili}. 

\section{Fireball Observation and Trajectory}

\begin{figure}
    \centering
    \includegraphics[width=1.0\textwidth]{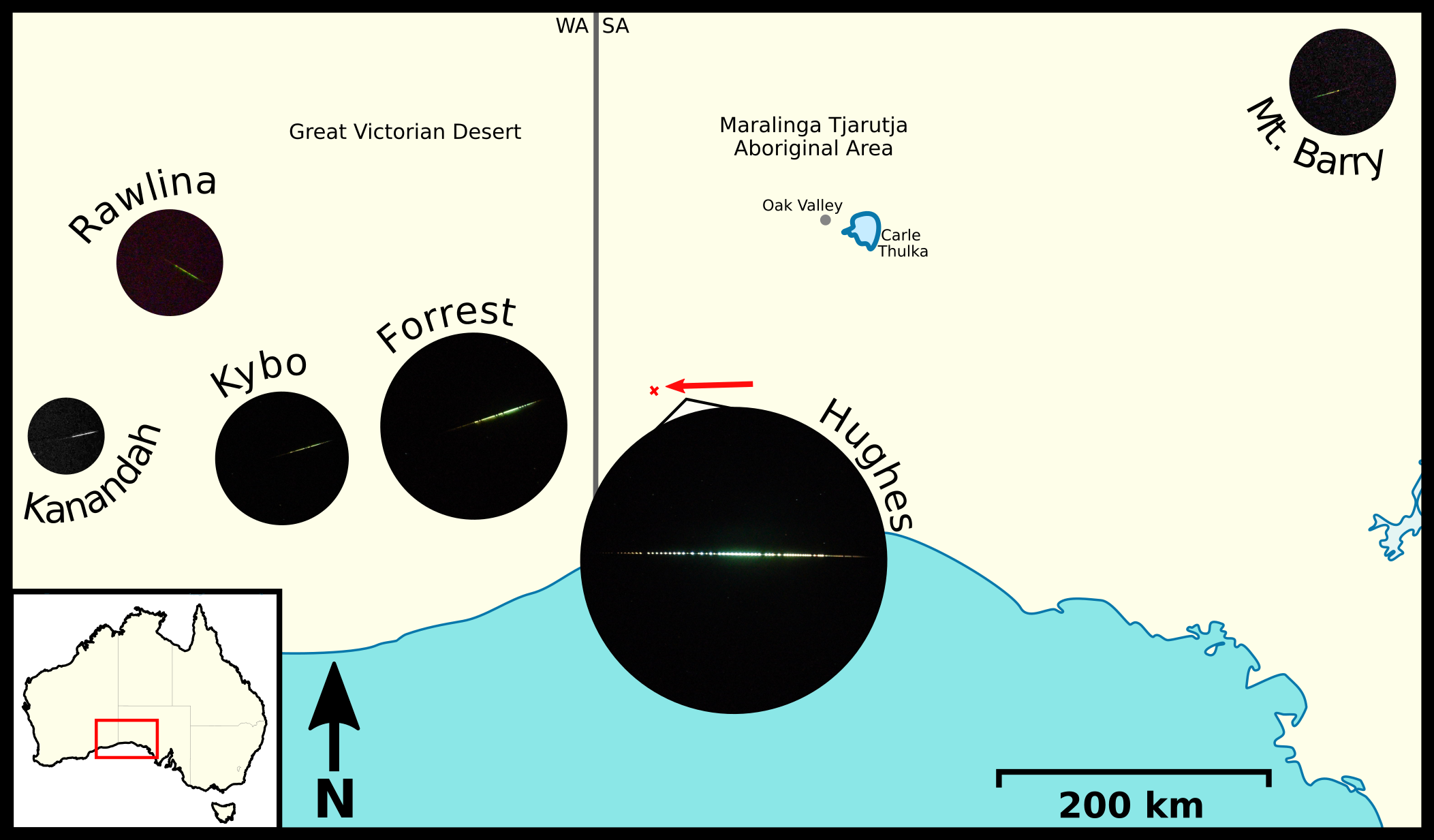}
    \caption{Cropped all-sky images of the fireball from the six DFN observatories. Images are of the same pixel scale, with the center of each image positioned at the observatory location on the map. Dashes encoded in the trajectories are an expression of the liquid crystal shutter modulation and provide both absolute and relative timing along each trajectory. Location of Arpu Kuilpu meteorite recovery is shown by the red cross, with the ground track of the trajectory as a red arrow. See supplementary material for full resolution images.}
    \label{fig:images_on_map}
\end{figure}

In total, six DFN camera systems detected the event internally referenced as \textit{DN190601\_01} (Tab. \ref{table:stations}), using the on-board algorithm of \citet{2020PASA...37....8T}. The location of these are mapped in Figure \ref{fig:images_on_map} along with the fireball as observed by each system.

The DFN captures long exposure images and obtains timing information from de Bruijn sequence dashes \citep{Howie2017Howbuildcontinental,Howie2017Submillisecondfireballtiming,Howie2019Absolutetimeencoding}. This methodology of encoding a de Bruijn sequence into the shutter frequency allows us to obtain absolute timing of the fireballs without the need of an additional subsystem -- drastically reducing the size, cost, and power requirements. 

Observing conditions during the fireball were ideal, enabling three cameras to observe from over 400\,km away. Encoded timing was acquired for five sites, but was not resolvable for DFNEXT051. The best convergence angle between the four observations was a reasonable 36$^{\circ}$. However, one camera, DFNEXT041 Hughes, was nearly directly underneath the fireball trajectory. This observation significantly helped to constrain the fit and was instrumental in recovering the small 42\,g stone meteorite. The observations from the six DFN fireball observatories provided 256 data points to fit the trajectory (Figure\ref{fig:images_on_map}). 

\begin{table}[!h]
	\caption{Locations of DFN Observatories that obtained photographic records of DN190601\_01, and nature of data obtained. Times are relative to first fireball observation at 09:53:04.326 UTC on the 1st of June 2019 (from the Hughes camera). Timing information was unable to be extracted from the observations made from DFNEXT051 - Kanandah.
	P: Photographic record (long-exposure high resolution image, see Sec. \ref{sec:astrometry}), V: lossless compressed digital video (30 frames per second, see Sec. \ref{sec:photometry}). Ranges are from when the meteoroid was at 65\,km altitude.
	}              %
	\label{table:stations}      %
	\center                                      %
	\resizebox{\textwidth}{!}{
	\begin{tabular}{l c c c | c | c c c}          %
		\hline\hline                        %
		 Observatory &&&&Instrument& range\tablenotemark{*} &  start time & end time \\
		name & latitude & longitude & altitude (m)  & record& (km) & observed & observed \\
		\hline      
		DFNEXT041 - Hughes & 30.652867 S & 129.700608 E & 144 &P, V& 70  & 0.0 & 5.0 \\ 
		DFNEXT027 - Forrest  &  30.858058 S & 128.115032 E & 166  &P&  190 & 0.67 & 4.47 \\  
		DFNEXT043 - Kybo &  31.024935 S & 126.591345 E & 176  &P&  332 & 0.87 & 4.35 \\ 
		DFNSMALL26 - Rawlina &  29.742252 S & 125.750340 E & 209  &P&  422  & 1.67 & 2.87 \\ 
		DFNEXT032 - Mount Barry & 28.516433 S & 134.886265 E & 169 & P & 538 & 1.74 & 3.64 \\
		DFNEXT051 - Kanandah & 30.901091 S & 124.884540 E & 195 & P & 495 & - & - \\
		\hline                                             %
	\end{tabular}}
		\tablenotetext{*}{distance from the observatory to the meteoroid at 65 km altitude}
\end{table}

\subsection{Astrometry from photographs}\label{sec:astrometry}

After time decoding and centroiding the fireball positions, we use the same method as described by \citet{2018M&PS...53.2212D} to convert the pixel positions in the image to true horizontal coordinates.
The calibration frame used for most observatories was taken at 2019-06-01T10:00:30, 7 minutes after the fireball, using $\simeq$ 1000-1500 reference stars for close cameras, and 2000+ using background subtraction on farther observatories (Mt Barry and Kanandah) in order to have maximize the number of star low on the horizon.
We propagate uncertainties all the way through, taking into account both pixel picking uncertainties and residuals on the star fit.
This yields formal astrometric uncertainties on the order of 1.5\arcmin (1$\sigma$) for most records.
Note that the same angular precision is achieved throughout the frame, thanks to a change of pixel pitch due to the stereographic lens projection, from 120\arcsec at the zenith, down to 70\arcsec at 5\degree on the horizon.

Astrometric data files for all cameras are available in the supplementary material, and documentation associated with these data are available at \url{https://dfn.gfo.rocks/data_documentation.html}

\newpage
\subsection{Photometry from video records}\label{sec:photometry}

The latest generation observatories of type \textit{DFNEXT} are equipped with a monochrome digital video camera (Point Grey/FLIR BFLY-U3-23S6M-C, Fujinon Fisheye 1:1.4/1.8mm). The effective field of view is an all-sky circle that fits in a square of $\simeq1080$ pixels. This addition in parallel to the still high-resolution imager was introduced to not only observe fainter meteors than in the photographs, provide observational coverage during the dead time (3\,s = 10\%) between long exposures, yield better photometry, and also observe during the daytime. These systems run a modified version of \textit{Freeture} \citep{2020A&A...644A..53C} to detect and save fireballs (code available at \url{https://github.com/desertfireballnetwork/freeture_DFN}). Frames are saved as 8-bit lossless compressed FITS files.

\begin{figure}[!h]
	\centering
	\includegraphics[height=\textheight]{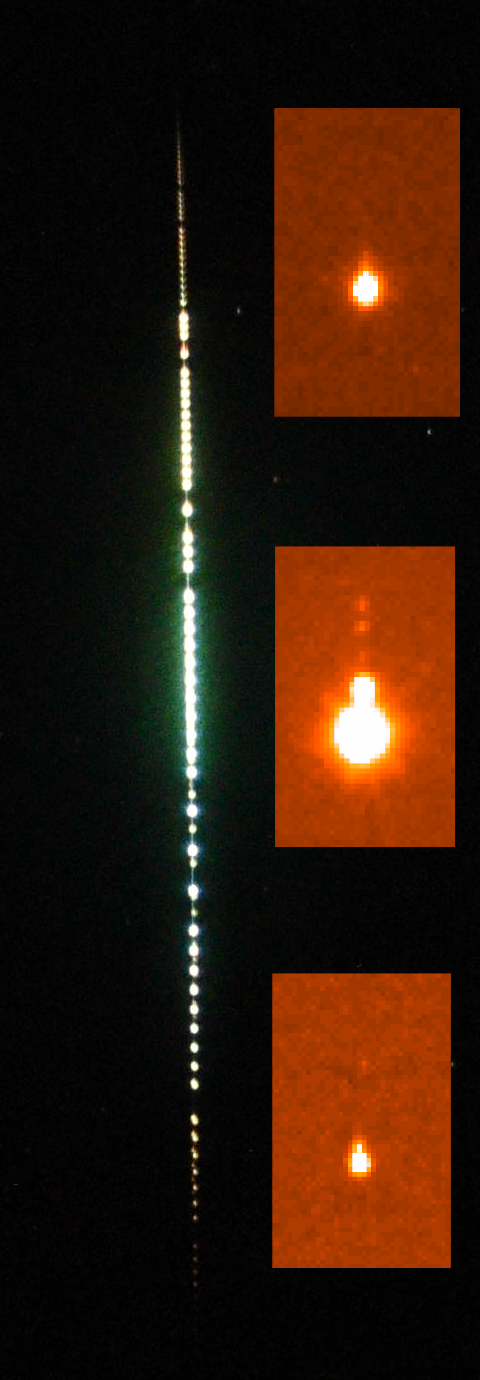}
	\caption{Still record from Hughes, along with 3 selected video frames from the same observatory. Fireball direction is top to bottom.}
	\label{fig:hughes_image}
\end{figure}

\begin{figure}[!h]
	\centering
	\includegraphics[width=\textwidth]{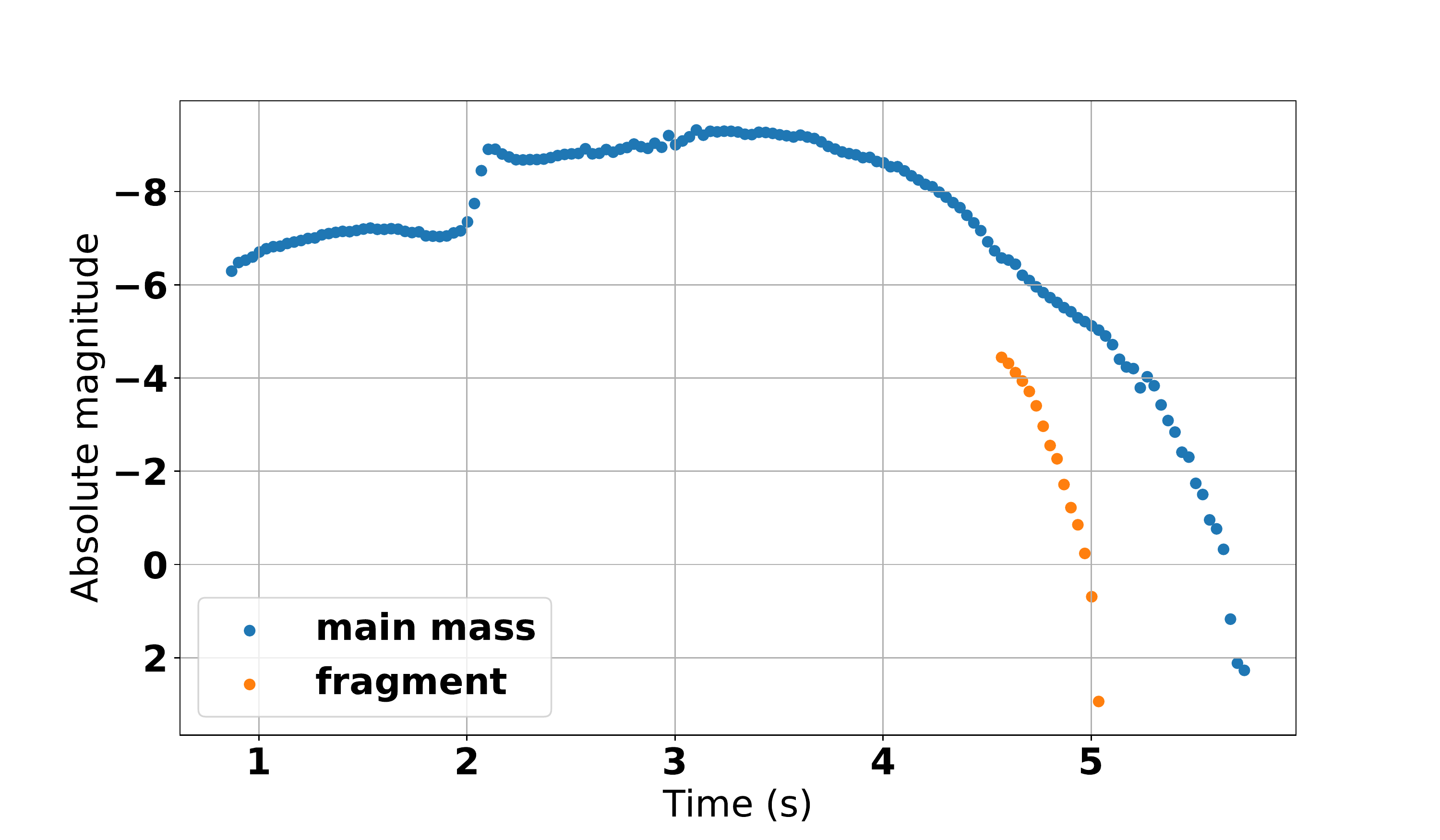}
	\caption{Absolute (distance corrected) light curve for the fireball. Time is relative to 2019-06-01T09:53:04Z.}
	\label{fig:light_curve}
\end{figure}

At the time of the event, the recording software deployed on the observatories was still in development: it was not recording long exposure calibration frames for astrometry and also failed to record the beginning of the fireball on \textit{DFNEXT041} (Tab. \ref{table:stations}).
Although the record was saturated in the brightest part of the fireball, we used a Point Spread Function (PSF) fitting photometry technique to estimate the true luminosity.
Compared to aperture photometry, PSF photometry mitigates the saturation issue to some extent, but still cannot guarantee the accuracy of the results.
To calibrate the records, we also perform PSF fitting on $\alpha$ Crucis to use it as a reference luminosity source ($V_{mag}$ = 1.25 after atmospheric extinction taken into account). Finally we use the triangulation solution (Sec. \ref{sec:modeling_and_orbit}) to determine the range of each observation in order to obtain the distance corrected (normalized to 100\,km distance) absolute magnitude (Figure \ref{fig:light_curve}). We also track the brightness of a secondary fragment for about half a second (Figure \ref{fig:light_curve}), $\simeq$2 magnitudes fainter than the main mass just after it broke apart. This trailing fragment is also visible in the still image (Figure \ref{fig:hughes_image}), in which a slight lateral deviation is visible. We note that in some frames, up to five fragments are visible in the video (main mass, aforementioned fragment, and three smaller ones; see the middle frame in Figure \ref{fig:hughes_image}).

\newpage
\section{Trajectory Modelling}\label{sec:modeling_and_orbit}

\subsection{Initial Triangulation} \label{sec:triang}
To determine a trajectory from the fireball observations, we utilized a modified version of the straight line least squares (SLLS) method of \citet{Borovivcka1990comparisontwomethods,Sansom2015novelapproachfireball}. The velocity and mass information is then extracted by using an extended Kalman smoother (EKS) \citep{Sansom2015novelapproachfireball}. Finally, camera observations are weighted according to their distance from the observed fireball, as the increased distance increases the cross-track uncertainties. 

The fitted trajectory was mostly constrained by the DFNEXT041 - Hughes, which was nearly directly underneath the trajectory (Tab. \ref{table:stations}). The fireball was seen for 5 seconds from an altitude of 86.4 km to 28.9 km. The meteoroid impacted the upper atmosphere at a modest angle ($\sim$50$^{\circ}$).
For further details about the fireball trajectory and dynamic information, please refer to Tab. \ref{tab:traj_sum}. 

The relatively featureless light curve for this fireball (Figure \ref{fig:light_curve}) indicates  few large fragmentation events. The increase in brightness at 2.07 seconds occurs at an atmospheric ram pressure of 0.11 MPa, while the event forming the largest secondary fragment at 4.6 seconds corresponds to pressures of 0.82 MPa. This is similar to the pressures experienced during fragmentation of other H5 chondrite falls such as the Grimsby, Ko\u{s}ice, and Ejby \citep{brown2011grimsbyfall,BorovivcKa2013Kosicemeteoritefall,spurny2017ejby}. Interestingly, the progenitor meteoroids for these falls span three orders of magnitude in mass, but each has initial fragmentations around $\sim$0.1\,MPa. This low fragmentation pressure range was proposed to be associated with reassembled and cemented material by \citep{borovivcka2020two}, meaning this feature may be constant within the $10^{0}$-$10^{3}$\,kg range for ordinary chondrites. 

\begin{figure}[!h]
	\centering
	\includegraphics[width=\textwidth]{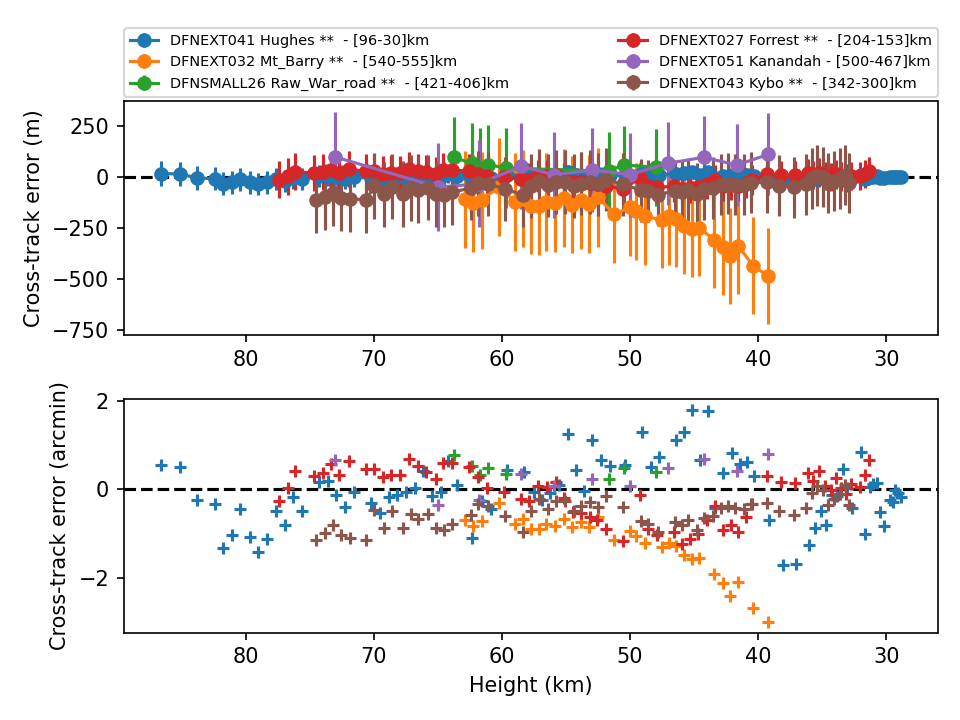}
	\caption{Cross-track residuals of the straight line least-squares fit to the trajectory from each viewpoint (bottom: in angular residuals. top: in distance projected on a perpendicular plane to the line of sight). The error bars displayed correspond to the 1-$\sigma$ uncertainties on the astrometry.
	Ranges given in the legend are from the camera to the [highest - lowest] trajectory point. %
	}
	\label{fig:cross_track_residuals}
\end{figure}

\subsection{Estimating initial and terminal masses}\label{sec:eks_grits}

\begin{figure}[!h]
	\centering
	\includegraphics[width=0.8\textwidth]{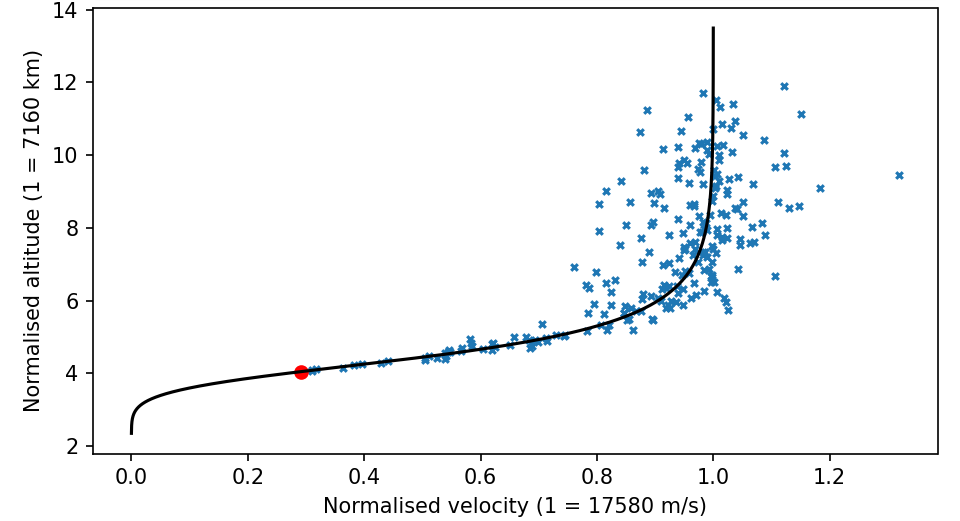}
	\caption{Trajectory data with velocities normalized to the velocity at the top of the atmosphere ($V_0$=17.58\,$\mbox{km s}^{-1}$; Tab. \ref{tab:traj_sum}) and altitudes normalized to the atmospheric scale height, $h_0 = 7160$\,m. The best fit to Equation 10 of \citet{Gritsevich2009Determinationparametersmeteor} results in $\alpha=36.16$ and $\beta=1.13$ and is shown by the black line.}
	\label{fig:alphabeta}
\end{figure}

We initially use the $\alpha$--$\beta$ methods of \citet{Gritsevich2012Consequencescollisionsnatural} and \citet{sansom2021metsoc} to determine a first approximation of final mass and velocity to initiate Bayesian filtering techniques. 
The dimensionless ballistic ($\alpha$) and mass loss ($\beta$) parameters calculated for this fireball are $\alpha=36.16$ 
and $\beta=1.13$ 
(Figure \ref{fig:alphabeta}). Assuming meteoroid properties, such as a brick-like shape (A=1.55; \citealt{Gritsevich2009Determinationparametersmeteor}), a bulk density of 3500\,kg\,m$^{-3}$, 
and a shape change parameter of 2/3 (see \citealt{2019ApJ...885..115S} and references therein), a terminal mass of 60 g is predicted, with an estimated initial mass of 1.4 kg. Assuming a spherical shape gives a minimum initial mass of 680 g, with a meteorite mass of 30 g. The fitted final velocity is 5.1 km/s.

The EKS filter predicts changes to the state (position, velocity, and mass) using the single body aerodynamic equations \citep{Sansom2015novelapproachfireball}. The state is initiated at $t_f=5.0$\,s, with state values taken from initial fitting above, with 10\% uncertainties in position and velocity, and 100\% in mass; for  velocity$=5.1\pm0.5\,\mbox{km s}^{-1}$ and mass$=0.06\pm 0.06$ kg.  Based on the deceleration of the object and a shape parameter once again of A=1.55, the final velocity was $=4.74\pm0.19\,\mbox{km s}^{-1}$, and the initial and final masses were estimated to be $2.2\pm0.1$\,kg and $0.04\pm0.01$\,kg respectively \citep{Sansom2015novelapproachfireball}. This is very consistent with the 42\,g meteorite sample recovered.

In order to get the best estimate of initial velocity for orbit calculation, we need to account for any variation in the trajectory from a straight line due to the atmosphere. This involves recalculating the trajectory and EKS modelling for the observations above 60 km altitude. In doing so, the initial velocity is calculated to be $=17.58\pm 0.04\,\mbox{km s}^{-1}$.

\begin{table}[!h]
	\centering
	\begin{tabular}{lll}
		\hline
	                                       	   &       Beginning           &        Terminal         \\
		\hline
		Time (UTC)                             & 2019-06-01T09:53:04.300   & 2019-06-01T09:53:09.300 \\
		Longitude (WGS84; $\degr$E)            & 130.132155                & 129.631139              \\
		Latitude (WGS84; $\degr$S)             & 30.610298                 & 30.623636               \\
		Height (km)                            & $86.44\pm0.03$            & $28.90\pm0.01$          \\
		Velocity ($\mbox{km s}^{-1}$)          & $17.58\pm0.04$            & $4.74\pm0.19$           \\
		Angle from local horizontal ($^\circ$) & 50.0                      & 49.7                    \\
		Apparent entry radiant (RA $^\circ$)   & 210.48$\pm$0.01   & --                      \\
		Apparent entry radiant (Dec $^\circ$)  & -22.08$\pm$0.01  & --                      \\
		Calculated mass\footnote{Assumptions listed in Section \ref{sec:eks_grits}} (kg) & $2.2\pm0.1$ & $0.04\pm0.01$ \\
		\hline
		Total duration               & 5.0 s                &                     \\
		Number of data points        & 256                  &                     \\
		Number of observing stations & 6 (5 with timing)                    &                     \\
		Recovery location (WGS84)    & 129.631274$^\circ$ E & 30.623689$^\circ$ S \\
		Recovered mass               & 42 g                 &                     \\
		\hline
	\end{tabular}
	\caption{Atmospheric trajectory information for event DN190601\_01. Apparent entry radiant and velocities are in an inertial, Earth-centered reference frame (ECI). }
	\label{tab:traj_sum}
\end{table}

\newpage
\section{Radiant and Heliocentric orbit determination}\label{sec:orb}
\begin{figure}
    \centering
    \includegraphics[width=0.8\textwidth]{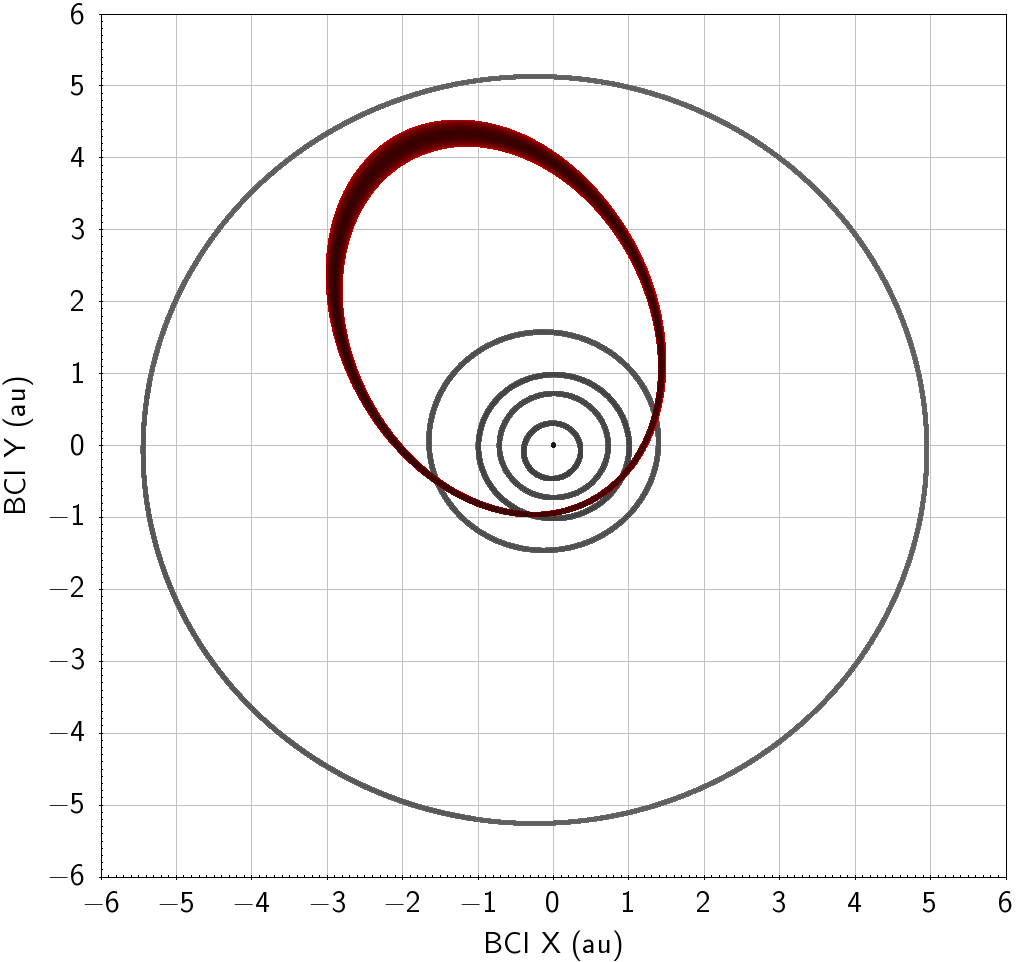}
    \caption{Orbit calculated from observations for the \textit{DN190601\_01} meteoroid (red). The orbit is displayed in the barycentric inertial reference frame (BCI), coinciding with that of the International Celestial Reference System (ICRS). Planets and the sun are displayed in gray.}
    \label{fig:orb}
\end{figure}

As stated in Section \ref{sec:modeling_and_orbit}, the straight-line trajectory may not be relied upon to provide representative entry vectors of the meteoroid from which to calculate orbits. Thus, the initial radiant and velocity were calculated using only the top of the trajectory ($>60$\,km) before major deceleration. This apparent radiant and velocity was used to calculate the pre-atmospheric orbit of the meteoroid using the integration method of \citet{JansenSturgeon2019Comparinganalyticalnumerical}. The Monte Carlo results are illustrated in Figure \ref{fig:orb}, with orbital values given in Tab. \ref{tab:orbit}. 

A Monte Carlo simulation was performed to characterize the orbital characteristics and history for Arpu Kuilpu. One thousand test particles were generated within the formal triangulation uncertainties assuming a Gaussian distribution (Tab. \ref{tab:orbit}). These particles were then propagated backward in time 10\,Myrs using the IAS15 integrator implemented through the Python-based REBOUND module, taking into account all relevant planetary perturbations (all planets, the Sun, and the Moon) \citep{2012rebound,2015_rebound_IAS15}. We found that $35.2\pm4.2\%$ of the particles were unstable over the previous 10\,kyrs, having numerous close encounters with Jupiter and terrestrial planets. As previously noted by \citet{tancredi2014criterion}, the unpredictability of the evolution of an object`s orbit on a 10\,kyrs timescale is a diagnostic feature of Jupiter-family comets. However, given our simulations, the Arpu Kuilpu meteoroid`s orbital evolution is more likely consistent with debris from the main asteroid belt. This is consistent with the findings of \citet{Shober2020Grazing,shober2020using,shober2021main}, which showed that nearly all of the centimeter- to meter-sized debris on Jupiter-family comet-like orbits are initially from the main belt. After $\sim20$\,kyrs, the orbital history of Arpu Kuilpu becomes too challenging to characterize. A significant number of close encounters for all the particles removes the ability to gain further information through backward integration. 

The orbit for Arpu Kuilpu (Tab. \ref{tab:orbit}) is very similar to Ko\u{s}ice, Hamburg, and Ejby, other H-chondrite falls recovered with orbits elements of a~$\sim$~2.7~au, e~$\sim$~0.65, inc~$<$~2$\degr$. \citet{granvik2018identification} determined the likely escape routes from the main-belt for Ko\u{s}ice and Ejby based on a debiased NEO orbital model \citep{granvik2018debiased}. They found that these meteorites' most likely transport mechanisms were the 3:1 or 5:2 mean motion resonances. Thus, Arpu Kuilpu should be expected to have a similar history. Based on the NEO model, the meteorites also had a very high probability of originating from the JFC population, but, as previously stated, this is expected for meteorites coming from comet-like orbits \citep{shober2020using,Shober2020Grazing,shober2021main}.

\begin{table}[!h]
    \centering
    \begin{tabular}{ccc}
    \hline 
    Epoch                                    & TDB              & 2019-06-01    \\
    Semi-major axis $a$                      & AU                & 2.75 $\pm$ 0.03            \\
    Eccentricity $e$                         &                   & 0.671 $\pm$ 0.003          \\
    Inclination $i$                          & \degr             & 2.03 $\pm$ 0.01            \\
    Argument of perihelion $\omega$          & \degr             & 43.25 $\pm$ 0.02           \\
    Longitude of the Ascending Node $\Omega$ & \degr             & 250.36 $\pm$ 0.01        \\
    Perihelion distance $q$                  & AU                & 0.90 $\pm$ 0.01          \\
    Aphelion distance  $Q$                   & AU                & 4.59 $\pm$ 0.05            \\
    Corrected radiant (RA) $\alpha _g$       & \degr             & 215.76 $\pm$ 0.01    \\
    Corrected radiant (Dec) $\delta _g$      & \degr             & -20.04 $\pm$ 0.01    \\
    Geocentric speed $V _g$                  & $\mbox{m s}^{-1}$ & 13280 $\pm$ 50             \\
    Tisserand`s parameter $T_J$              &                   & 2.97 $\pm$ 0.02            \\ 
    \hline 
    \end{tabular}
    \caption{Pre-encounter orbital parameters expressed in the heliocentric ecliptic frame (\textit{J2000}) and associated $1\sigma$ formal uncertainties.}
    \label{tab:orbit}
\end{table}

\section{Wind modeling and Darkflight}

\begin{figure}
    \centering
    \includegraphics[width=0.8\textwidth]{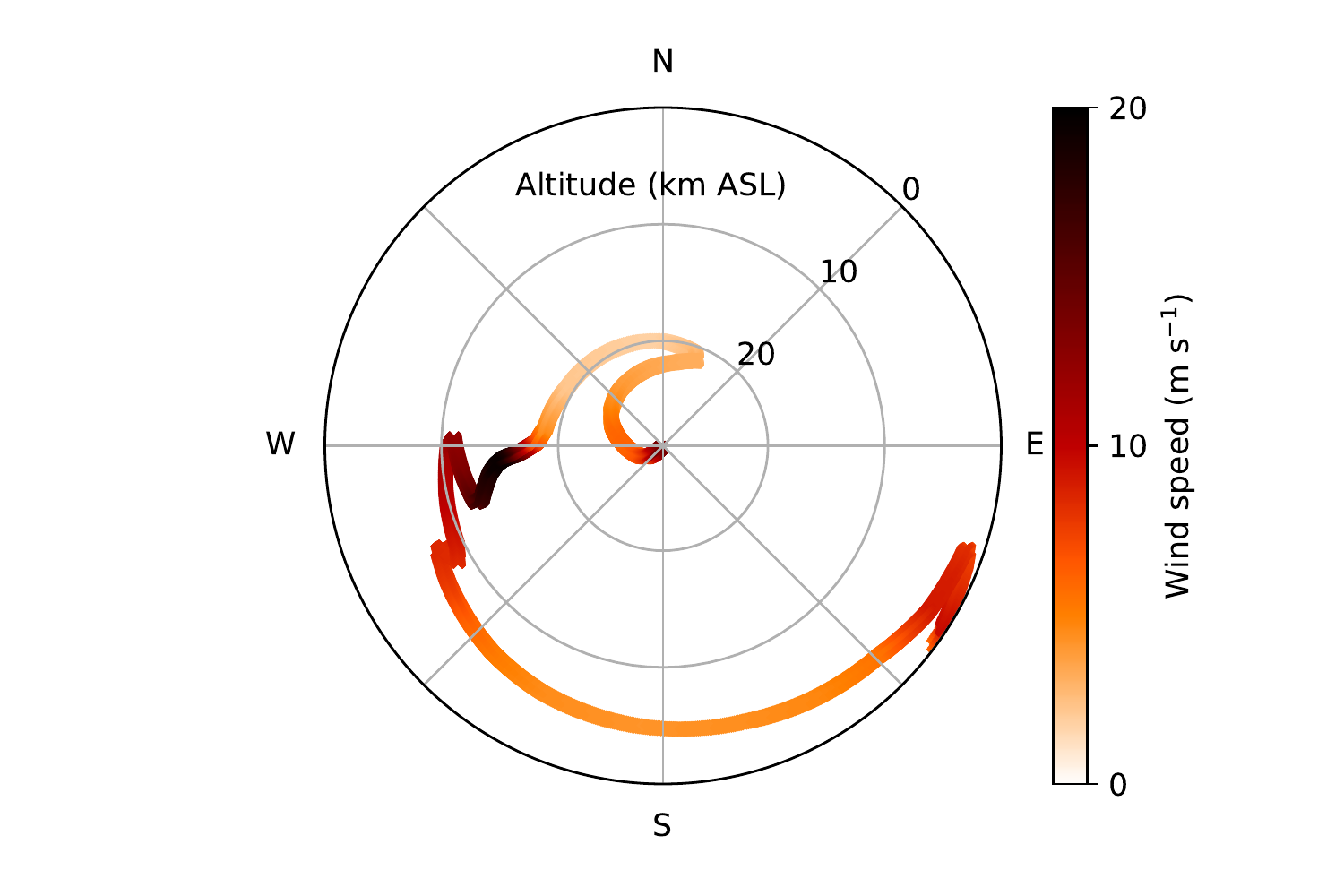}
    \caption{Wind model (speed and direction for a given altitude), extracted as a vertical profile at the coordinates of the lowest visible bright flight measurement. Model integration started at 2019-06-01T00:00. The winds affecting this fall were moderate, as the maximum wind encountered by the meteoroid during the dark flight was $\sim$19\,m\,s$^{-1}$ coming from the West, at around 15 km altitude.}
    \label{fig:Wind}
\end{figure}

We use the Weather Research and Forecasting (WRF) model version 4.0 with dynamic solver ARW (Advanced Research WRF)  \citep{Skamarock2019descriptionadvancedresearch} for atmospheric situation numerical modeling.
The weather profile (Figure \ref{fig:Wind}) includes wind speed and direction at heights ranging up to 30 km, fully covering the dark flight altitudes ($<28.9$\,km).
The dark flight trajectory of the meteorite was substantially affected by the atmospheric winds, dominated by the subtropical jet stream at altitudes approx. 10-15km. However, the winds during the Arpu Kuilpu fall were not as strong as in the case of Murrili \citep{sansom2020murrili} or Dingle Dell \cite{Devillepoix2018DingleDellmeteorite}. 
From the 3D data produced by the WRF/ARW modeling software, we extract weather parameters relevant for the dark flight modeling (wind speed, wind direction, pressure, temperature, and relative humidity). We then interpolate these conditions in time and 3D coordinates during darkflight modeling, following the meteorite fall trajectory at exact time and space \citep{2021arXiv210804397T}.

\begin{figure}
    \centering
    \includegraphics[width=1.\textwidth]{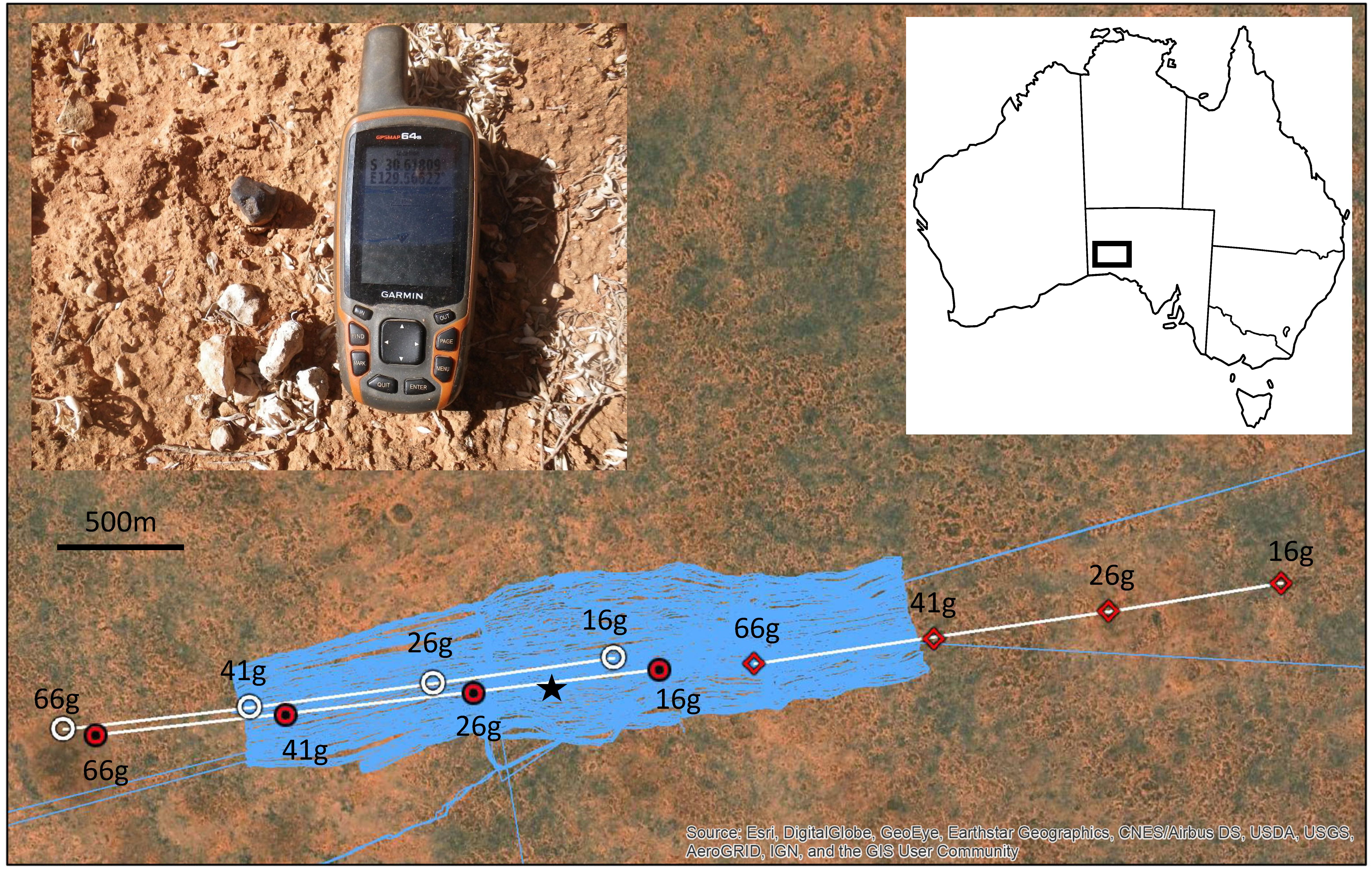}
    \caption{The fall area lies within the South Australia area of the Nullarbor. Fall lines are shown with predicted masses for three scenarios; a hypothetical chondritic sphere (red circle) and chondritic cylinder (red diamond) using the preferred wind model, and a chondritic sphere with an alternative wind model (white circle). Meteorite recovery position shown as filled black star, with an inset image of the meteorite as found with GPS unit for scale. Blue tracks show the areas searched by the team. The location of the recovered meteorite is about 30\,m from the chondritic sphere fall line, but corresponding to a mass of about 20\,g, compared to a recovered mass of 42\,g.}
    \label{fig:fall_line_draft}
\end{figure}

From this wind profile and the mass and shape estimates above, we generate fall position predictions on the ground, showing hypothetical locations for a range of possible masses. Using a range of possible masses with all else equal, one can construct a ground fall line, to guide searching. However, this line is a construct of scenarios, it does not imply that pieces are present all along the line. If one further considers all possible scenarios, such as different wind model predictions and different shapes as well as possible masses, a heat map of possible fall positions can be made, see for example \citep{moilanen_determination_2021}. However, in Figure \ref{fig:fall_line_draft} we plot lines for only three possible scenarios for clarity (white lines), as detailed in the caption. Other possible scenarios are easily generated (see supplemental data). The environmental conditions at this fall were generally quite stable, and this can be seen in Figure \ref{fig:fall_line_draft}, where the fall lines for different wind models, but for the same hypothetical shape are relatively close together. 

\section{Search and recovery}
Although the expected main mass was quite small, this particular fall was prioritized for searching thanks to several factors:
\begin{itemize}
    \item Very clean data, both for trajectory and dynamics, thanks to a close-by viewpoint (Hughes station). This gave us confidence the search area would be very well constrained, and the mass would be well bracketed.
    \item Lack of recent rain and therefore vegetation likely sparse in the area.
    \item Relatively low-risk access to the site.
    \item Interesting JFC-like orbit with larger semi-major axis.
\end{itemize}

A team of six people carried out a searching expedition for two weeks from 17 July 2019 onward.
On-site, the searching conditions were excellent, with most vegetation being sparse low bluebush and saltbush, with little to no grass, and sizable barren clay pan areas. The meteorite was recovered on the second day of searching, but searching continued for the entire trip. As a searching strategy, the spacing between party members was 2-4~m, even though the seeing was excellent, based on the probable scenario of a relatively small meteorite. The meteorite was spotted about 1~m away from the searcher.

The meteorite was on the ground for approximately six weeks: The nearest weather station is Forrest airport, in Western Australia, and during the period of 1 June to 20 July, records show $8.4$ mm of rain. Nullarbor rainfall is relatively localized, mainly resulting from thunderstorms, so more significant rainfall at the fall site cannot be ruled out. During winter, it is also not unusual for ground fog and dew to occur in the Nullarbor.

\begin{figure}
    \centering
    \includegraphics[width=\textwidth]{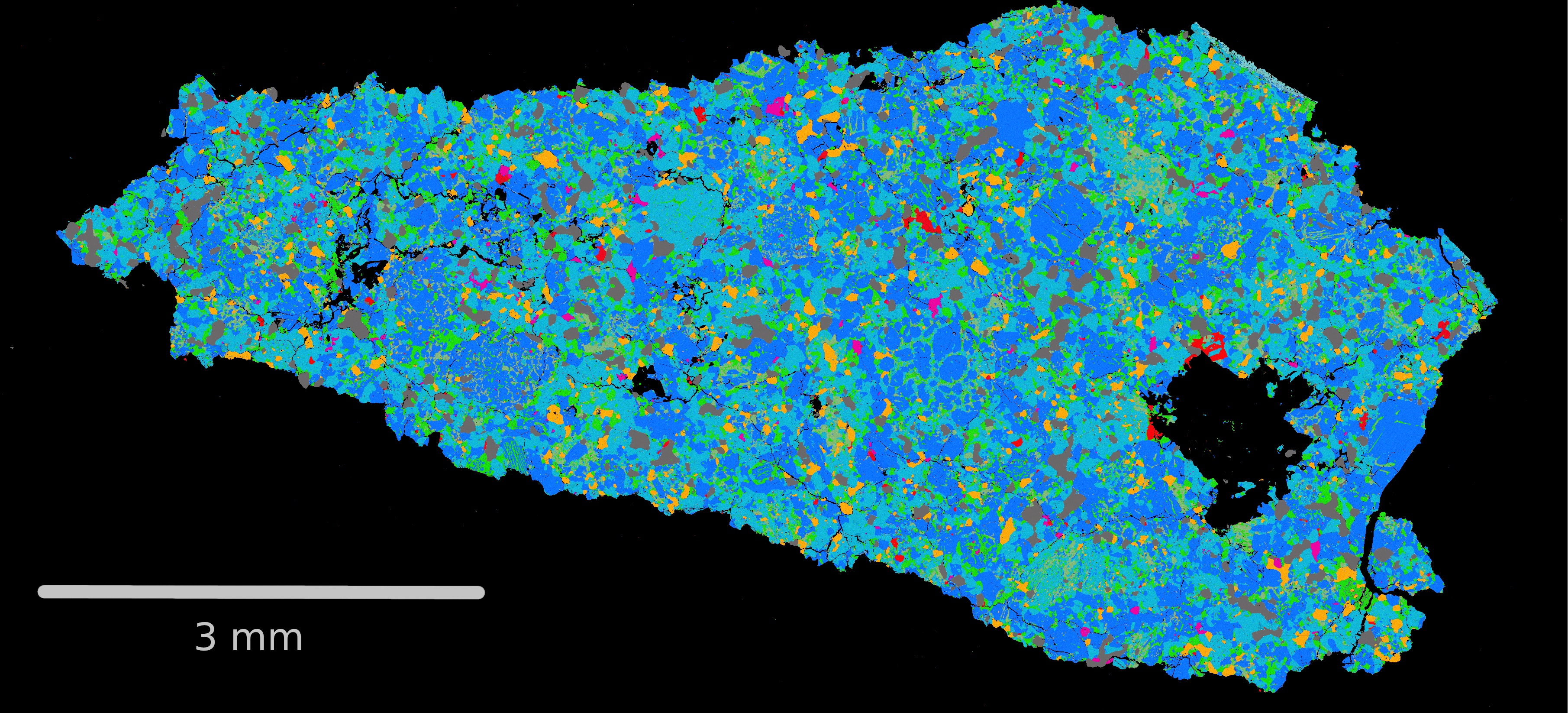}
    \caption{Mineral map of Arpu Kuilpu section constructed from Fe, Ni, Mg, Ca, S, Si, Cr element maps obtained by scanning electron microscopy (SEM) with energy dispersive X-ray spectrometry (EDS). blue = olivine; teal = orthopyroxene; green/brown = clinopyroxene; yellow/green = plagioclase; orange = troilite; red = phosphate; pink = chromite; gray = FeNi metal}
    \label{fig:SEM}
\end{figure}

\subsection{The meteorite sample}

The recovered meteorite sample consists of a single, rounded stone $\sim$1-1.5~cm in diameter, with no noticeable terrestrial alteration. A portion of the meteorite was set in epoxy, polished, and carbon coated to enable classification via Scanning Electron Microscope (SEM) and Electron Probe Micro-Analysis (EPMA) (Figure~\ref{fig:SEM}). Arpu Kuilpu's mineralogy and texture is typical of equilibrated ordinary chondrites, containing relict chondrules, metal, and troilite grains suspended in a recrystallized silicate matrix. The relative abundances of plagioclase, olivine, pyroxene, metal, and troilite within the sample, combined with the chemical compositions of its olivines, pyroxenes, and chromites confirm that Arpu Kuiplu is an H5 chondrite. 

Cosmogenic radionuclide concentrations have been analyzed by means of non-destructive high purity germanium (HPGe) gamma spectroscopy. The counting efficiencies have been calculated using thoroughly tested Monte Carlo codes. One specimen of Arpu Kuilpu was measured in the underground laboratories at the Laboratori Nazionali del Gran Sasso (LNGS) \citep{arpesella1994low,laubenstein2017screening} for 40.63 days (183 days after the fall). Very low activity of $^{60}$Co ((0.6 $\pm$ 0.3) dpm/kg) suggests that the pre-atmospheric size of the Arpu Kuilpu meteoroid was rather small and no significant production of secondary thermal neutrons took place within the meteoroid during its recent cosmic ray exposure in space. The measured $^{26}$Al activity is consistent with that expected for a small-size H chondrite \citep{bhandari1989torino,bonino2001solar,leya2009cosmogenic}.

When we compare the radionuclide concentrations with cosmic ray production estimations for $^{26}$Al \citep{leya2009cosmogenic}, $^{60}$Co \citep{eberhardt1961neutrons,spergel1986cosmogenic}, $^{54}$Mn \citep{kohman1968nuclide}, and $^{22}$Na \citep{bhandari1993depth}, and assume the specimen is from the central part, the best agreement is obtained for radii of $<$\,10 cm, $<$\,20 cm, $<$\,5 cm and 5-8~cm, respectively. Combining all results of these radionuclides, we infer for a roughly spherical meteoroid with of about 5 to 10 cm in radius. This estimate corresponds to an initial mass of $\sim$1.8-14.7~kg. The dynamic mass determined using the deceleration of the meteoroid in the atmosphere (2.2~kg; Tab.~\ref{tab:traj_sum}) favors the lower end of this estimate. 
Therefore, the true initial mass of the meteoroid was likely $\sim$2\,kg.
The small size of the progenitor body and minimal fragmentation observed indicates it was a single boulder with negligible macroscopic fractures.

For further information, please refer to the Meteoritical Bulletin Database\footnote{\url{https://www.lpi.usra.edu/meteor/metbull.php?code=74013}}. A forthcoming paper will more thoroughly discuss the sample analysis. 

\section{Conclusions}
Arpu Kuilpu is the fifth meteorite to be recovered by the DFN in Australia. The sample was found within close proximity to two of the three predicted fall lines, varying the shape and wind model. The predicted final mass (40$\pm$10\,g) is very close to the recovered stone`s mass (42\,g).
With an initial mass of $\sim$\,2\,kg, the meteoroid is by far the smallest to ever survive entry and be recovered as a meteorite (the Cavezzo meteorite being second smallest at $\simeq 3.5 kg$ \citep{2021MNRAS.501.1215G}).
The diminutive size of the progenitor body also demonstrates the robustness of our data reduction pipeline. Additionally, it supports the DFN`s original logic that desert regions are prime locations for fireball networks, where the likelihood of a successful meteorite recovery is greatest \citep{Bland2012AustralianDesertFireball}. Arpu Kuilpu is an H5 ordinary chondrite from the outer main belt (a $\sim$ 2.75~au). The orbital evolution over the previous 10\,kyrs is consistent with asteroidal debris, despite the proximity to a comet-like Tisserand`s parameter ($T_{J}$=2.97). This is consistent with the results of \citep{shober2021main}, which demonstrated that nearly all sporadic comet-like debris seen by fireball networks are asteroidal in origin. Given the orbit, the most likely escape routes from the main belt are the 3:1 or 5:2 mean motion resonances \citep{granvik2018debiased,granvik2018identification}.

\section*{Supplementary material}

Supplementary material has been uploaded as a \textit{Zenodo} record at \url{http://doi.org/10.5281/zenodo.5775303}.
It contains fireball images (including calibration data), astrometry tables, the trajectory, wind profiles, as well as meteorite searching records.

\section*{Acknowledgments}
The authors would like to thank the volunteers who assisted on the meteorite search trip: Carlo Mungioli, Zdenak Martelli, and Aurelio Martelli. We would also like to thank the traditional owners of the land, the Maralinga Tjarutja people, who gave us permission to search and helped us name the meteorite recovered.

This work was funded by the Australian Research Council as part of the Australian Discovery Project scheme (DP170102529), and receives institutional support from Curtin University. Data reduction is supported by resources provided by the Pawsey Supercomputing Centre with funding from the Australian Government and the Government of Western Australia. The DFN data reduction pipeline makes intensive use of Astropy, a community-developed core Python package for Astronomy \citep{AstropyCollaboration2013AstropycommunityPython}. This research made use of TOPCAT for visualization and figures \citep{taylor2005topcat}. Simulations in this paper made use of the REBOUND code which can be downloaded freely at http://github.com/hannorein/REBOUND \citep{2012rebound}. The authors declare no competing interests.

\bibliography{biblio}
\bibliographystyle{aasjournal}

\end{document}